\documentclass[superscriptaddress,preprintnumbers,amssymb,nofootinbib,twocolumn]{revtex4}

\usepackage{subfig}
\usepackage{bm}
\usepackage{esvect}
\usepackage{verbatim}
\usepackage{multirow}

\usepackage[left]{lineno}
\usepackage{blindtext}
\pdfoutput=1
\usepackage[T1]{fontenc}
\usepackage{graphicx}
\usepackage{epsfig}
\usepackage{amsmath}
\usepackage{amsfonts}
\usepackage{amssymb}
\usepackage{braket}
\usepackage{cancel}
\usepackage{pstricks}
\usepackage{color}
\usepackage{feynmf}
\usepackage[normalem]{ulem}
\usepackage{epstopdf}
\usepackage{soul}
\usepackage{cancel}
\usepackage{appendix}

\definecolor{ao(english)}{rgb}{0.0, 0.5, 0.0}

\def\tp{{\cal O}_{\tt CP}}

\usepackage{braket}
\usepackage{cancel}
\usepackage{setspace}
\usepackage{pstricks}
\usepackage{xcolor}
\usepackage{float}
\usepackage{mathtools}
\usepackage{multirow}
\usepackage{booktabs}
\usepackage{enumitem}
\providecommand{\tabularnewline}{\\}

\def\gsim{\lower0.5ex\hbox{$\:\buildrel >\over\sim\:$}}
\def\lsim{\lower0.5ex\hbox{$\:\buildrel <\over\sim\:$}}

\begin{document}

\title{Generic tests of CP-violation in high-$p_\text{T}$ multi-lepton signals at the LHC and beyond}
\author{Yoav Afik}
\email{yoavafik@gmail.com}
\affiliation{Enrico Fermi Institute, University of Chicago, Chicago, IL 60637, USA}
\affiliation{Experimental Physics Department, CERN, 1211 Geneva, Switzerland}
\author{Shaouly Bar-Shalom}
\email{shaouly@physics.technion.ac.il}
\affiliation{Physics Department, Technion--Institute of Technology, Haifa 3200003, Israel}
\author{Kuntal Pal}
\email{kpal002@ucr.edu}
\affiliation{Physics Department, University of California, Riverside, CA 92521, USA}
\author{Amarjit Soni}
\email{adlersoni@gmail.com}
\affiliation{Physics Department, Brookhaven National Laboratory, Upton, NY 11973, USA}
\author{Jose Wudka}
\email{jose.wudka@ucr.edu}
\affiliation{Physics Department, University of California, Riverside, CA 92521, USA}

\date{\today}

\begin{abstract}
We introduce a modification to the standard expression for tree-level CP-violation in scattering processes at the LHC, which is important when the initial state in not self-conjugate. Based on that, we propose a generic and model-independent search strategy for probing tree-level CP-violation in inclusive multi-lepton signals. We then use TeV-scale 4-fermion operators of the form $tu\ell\ell$ and $tc \ell \ell$ with complex Wilson coefficients as an illustrative example and show that it may generate ${\cal O}(10\%)$  CP asymmetries that should be accessible at the LHC with an integrated luminosity of ${\cal O}(1000)$~fb$^{-1}$.

\end{abstract}

\maketitle
\flushbottom

\newpage

The nature of CP-violation (CPV), which is closely related to the flavor structure, is one of  the major unresolved problems in particle physics. Indeed, the 
search for new CP-violating sources, beyond the standard model (SM), may be the key to a deeper 
understanding of particle physics and the evolution of the universe,
since CPV has far-reaching implications for cosmology \cite{Sakharov,Kuzmin:1985mm,Branco:2011zb}; in particular, the strength of CPV effects in the SM is insufficient to explain the observed baryon asymmetry of the universe (BAU), see e.g., \cite{Rubakov:1996vz,Bernreuther:2002uj,Canetti:2012zc}. It is, for these reasons, that the search for CPV beyond the SM is a very important component of the on-going effort for unveiling the physics that underlies the SM, even if the latter has already been observed.

In this paper we re-examine the formulation of tree-level CP-violating effects in scattering processes at 
the LHC, introducing a new term to the "master" CPV expression, which
properly identifies  the genuine CP violating signal and also takes into account "fake" CP-violating effects that arise when the initial state in not self-conjugate. We then        
present a generic test of CPV  in scattering processes, which is potentially sensitive to a wide variety of underlying new physics (NP) scenarios. We are particularly interested in CPV in the inclusive tri-lepton and four-lepton signals:
\begin{eqnarray}
    && pp \to \ell^{\prime -} \ell^+ \ell^- + X_3 \label{3l_fs} ~ \\
    && pp \to \ell^{\prime +} \ell^- \ell^+ + \bar{X}_3 \label{3l_fs_bar}~, \\
    && pp \to \ell^{\prime +} \ell^{\prime -} \ell^+ \ell^- + X_4 \label{4l_fs}    
\end{eqnarray}
where $\ell,\ell^\prime=e,\mu,\tau$ (preferably $\ell \neq \ell^\prime$, see below) and $X_3$, $\bar{X}_3$ and $X_4$ contain in general jets and missing energy. These include the $e^\pm \mu^+ \mu^-$ and $\mu^\pm e^+ e^-$ final states for $\ell,\ell^\prime=e,\mu$ and similarly for the pairs $\ell,\ell^\prime=e,\tau$ and $\ell,\ell^\prime=\mu,\tau$, as well as the 3-flavor final state $pp \to e \mu \tau +X$. As an example, we will consider below CPV in the $e^\pm \mu^+ \mu^-$ tri-lepton signals, but it should be clear that it is equally important to search for CPV in multi-leptons final states with as many different combination of flavors as possible.

Multi-lepton final states with high transverse momentum 
($p_T$) particles have been extensively studied at the LHC, both in measurements of SM processes and in searches for NP. However, searches for CP-asymmetries in such processes have been limited \cite{CPV_highpT1,CPV_highpT2,CMS:2022dbt}. 
Indeed, high-$p_\mathrm{T}$ charged leptons  are rather easily identifiable  objects with excellent resolution and are, therefore, very useful probes of generic NP at the LHC~\cite{ATLAS:2021wob,Sirunyan:2020tqm,CMS:2022dbt};\footnote{We note that final states involving the $\tau$ have, in general, a lower experimental detection efficiency and are, therefore, expected to be less effective for our study.} they are sensitive to many types of well-motivated underlying NP phenomena, such as lepton-flavor violation, lepton-universality violation, lepton-number violation~\cite{LFU-our,ourttllpaper,ourLQEFT,ATLAS_3l,CMS_heavy_lepton_3l,ATLAS_heavyN_1,theory_heavyN_1,theory_heavyN_2,theory_heavyN_3,theory_heavyN_4,theory_heavyN_5,ATLAS_doubely_C_Higgs,ATLAS:2022pbd} and CPV, which is the subject studied in this paper.
These multi-leptons signatures are also useful channels for searching for NP in top-quark systems and this has led to experimental searches e.g., in single-top and top-pair production processes $pp \to t\bar t V,~t \bar tH,~tV$~\cite{ttV_1,ttV_2,ttH_1,ttH_2} as well as in 4-top production $pp \to tt \bar t \bar t$~\cite{tttt_1,tttt_2} and searches for flavor-changing (FC) top physics ~\cite{tull-our,tZprime,FC_top1,FC_top2,ATLAS:FCtop1,ATLAS:FCtop2,ATLAS:FCtop3,ATLAS:FCtop4,ATLAS:FCtop5,ATLAS:FCtop6}.

The available momenta of the charged leptons in the final state of these multi-lepton signals allow a straightforward construction of CPV observables in the laboratory frame, as will be shown below. We note, though, that special care is needed for CPV tests at $pp$ colliders, where the initial state is not self-conjugate and the parton distribution functions (PDF's) of the incoming partons may, therefore, have an asymmetric structure.
This will be discussed below.

It should be emphasized that a sizeable, say ${\cal O}(\gtrsim 1\%)$ manifestation of CPV in multi-leptons events of the type \eqref{3l_fs}, \eqref{3l_fs_bar} or \eqref{4l_fs} will be  strong evidence for NP, since the CP-odd CKM-phase of the SM (which 
is responsible for CPV in the quark sector and has been measured \cite{PDG_CPV}) is expected to yield negligible CP-violating effects in these processes, as it can only arise from EW processes at higher loop orders \cite{ourreview}.\footnote{Note that, although B-decays can be a source of sizeable CP-asymmetries in the SM, 
their effect in final states with 3 or more leptons  of the type considered here is negligible.
Moreover, leptons from B-meson decays can be strongly suppressed using a number of kinematic properties.}
Furthermore, new CP-violating effects in leptonic systems may shed light on Leptogenesis, where the BAU is generated from a lepton asymmetry via a decay of a heavy neutral lepton~\cite{leptogenesis1,leptogenesis2}. 

Finally, we recall that in the last several years a few $\sigma$ deviations from the SM in $B$-decays \cite{BA1,BA2,BA3,BA4,BA5,BA6,BA7,BA8,BA9,BA10,BA11,BA12,BA13,BA14,BA15,BA16,BA17} as well as in the muon $g-2$ \cite{BNL:g-2,gm2-recent,Muong-2:2023cdq} have been measured, indicating a possible need for NP. The CPV searches in collider physics that are being suggested here are then especially timely since CP is not a symmetry of nature and, on general grounds, one expects new physics to entail beyond the SM CP-odd phase(s) \cite{Nir:1999mg,ourreview}.

Potential large tree-level CP-asymmetries at the LHC in the tri- and four-lepton production processes  \eqref{3l_fs}, \eqref{3l_fs_bar} and \eqref{4l_fs} can be searched for, using the following triple products (TP) of the lepton momenta ($\ell \neq \ell^\prime$):\footnote{The TP's in \eqref{TP1} are defined in the laboratory frame and we expect that systematic uncertainties in the reconstruction of the momenta involved will be smaller then e.g., the case where the momenta are defined in a rest frame of some particle(s). Also, the kinematical cuts on the leptons involved should be CP-symmetric, e.g., same $p_T$ cuts should be applied to all leptons.}$^{,}$\footnote{Useful TP correlations for CP studies in scattering processes at the LHC, which involve leptons with jets momenta, e.g., in $t \bar t$ and vector-bosons production, have been also suggested in~\cite{TP1,TP2,TP3,TP4,TP5,TP6}.}
\begin{eqnarray}
       \tp &=& \Vec{p}_{\ell^{\prime -}} \cdot \left( \Vec{p}_{\ell^+} \times \Vec{p}_{\ell^-} \right)  ~, \nonumber \\
    \overline{\tp} &=& \Vec{p}_{\ell^{\prime +}} \cdot \left( \Vec{p}_{\ell^-} \times \Vec{p}_{\ell^+} \right) \label{TP1}~,
\end{eqnarray}
which are odd under $P$ and under naive time reversal ($T_N$): 
time $\to$ -time. Under $C$ and $CP$ 
they transform as:
\begin{equation}
\begin{array}{ll}
       C(\tp) =  +\overline{\tp} \, ~, & C(\overline{\tp}) =  + \tp  \,~, \cr
       CP(\tp) =  -\overline{\tp} \,~, & CP(\overline{\tp}) =  - \tp  \,~. 
\end{array}
\label{CP-prop}
\end{equation}

Thus, to measure a nonzero TP
correlation effect for the $\tp$'s defined in \eqref{TP1}, the following 
$T_N$-odd (and also $P$-violating) asymmetries can be constructed: 
\begin{eqnarray}
A_T &\equiv& \frac{N\left( \tp >0 \right) - N\left( \tp < 0 \right)}{N\left( \tp >0 \right) + N\left( \tp < 0 \right)} \label{AT1} ~, \\
\bar{A}_T &\equiv& \frac{N\left( -\overline{\tp} >0 \right) - N\left( -\overline{\tp} < 0 \right)}{N\left( -\overline{\tp} >0 \right) + N\left( -\overline{\tp} < 0 \right)} \label{AT2} ~,
\end{eqnarray}
where $N\left( \tp >0 \right)$ is the number of events for which ${\rm sign}(\tp) > 0$ is measured etc. 

As will be shown below, a measurement of $A_T \neq 0$ and/or $\bar{A}_T \neq 0$
may indicate the presence of CPV (CP-odd phase(s)), but may also be a signal of some strong or generic CP-even phase, e.g., from final state interactions (FSI)~\cite{ourreview,Bar-Shalom:1995quw,Bar-Shalom:1997vih}, even if the underlying dynamics that drives the processes under consideration is CP-conserving. 
Therefore, in order to better isolate the pure CPV effect, we use  the following observable, sensitive to CPV: 
\begin{eqnarray}
    A_{CP} &=& \frac{\left(A_T - \bar{A}_T \right)}{2} ~.  \label{ACP1}
\end{eqnarray}

$A_{CP}$ may, in fact, also be ``contaminated'' by CP-even phases when the initial state is not CP-symmetric, as can be the case at the LHC or at $pp$ colliders, in general.
To see this, let us consider the underlying (hard) processes of the tri-lepton signals of \eqref{3l_fs} and \eqref{3l_fs_bar} (the discussion below applies similarly to the four-lepton signals of \eqref{4l_fs}): $a b \to \ell^{\prime -} \ell^+ \ell^- + X$ and $\bar{a} \bar{b} \to \ell^{\prime +} \ell^- \ell^+ + \bar X$. We assume for simplicity that there are only 2 interfering amplitudes that contribute to these processes as follows (CPV requires at least two amplitudes with different phases for any given process):
\begin{eqnarray}
    {\cal{M}}_{a b \to \ell^{\prime -} \ell^+ \ell^-} &=& M_1 e^{i \left(\phi_1 + \delta_1 \right) } + M_2 e^{i\left( \phi_2 + \delta_2 \right) }  \label{M} ~,
\end{eqnarray}
where we have factored out the CP-odd phases, $\phi_{1,2}$, and CP-even phases $\delta_{1,2}$. The latter typically arise from FSI at higher loop orders.
Also, $M_i$ can be
complex in general (as in our case below) 
and the amplitude for the charge-conjugate (CC) channel ($\bar{a} \bar{b} \to \ell^{\prime +} \ell^- \ell^+$) is obtained from
\eqref{M} by changing the sign of the CP-odd phases $\phi_i \to -\phi_i$ and replacing $M_i \to M_i^\star$.

The corresponding (hard) differential cross-sections can then be schematically written as: 
\begin{eqnarray}
    d \hat\sigma &=& W+ U \cdot \cos(\Delta\delta + \Delta\phi) + V \cdot \tp \cdot \sin(\Delta\delta + \Delta\phi) ~, \nonumber \\ 
     \label{sig}
\end{eqnarray}
and $d \bar{\hat\sigma}
=d \hat\sigma(\Delta\phi \to - \Delta\phi,\tp \to \overline{\tp})$ for the CC channel,
where $\Delta\phi = \phi_1 -\phi_2$, $\Delta \delta=\delta_1 - \delta_2$, $W \propto |M_1|^2, |M_2|^2$, $U \propto {\rm Re}\left(M_1 M_2^\dagger \right)$ and the 3rd term in \eqref{sig} arises from ${\rm Im}\left(M_1 M_2^\dagger \right) \propto \tp$ and is where the tree-level CPV resides, i.e., when $\Delta\delta=0$. 

We then find for $A_T$ and $\bar{A}_T$ in \eqref{AT1} and \eqref{AT2}: 
\begin{eqnarray}
A_T = {\cal I}_{ab} \sin(\Delta\delta + \Delta\phi) ,~
\bar{A}_T = {\cal I}_{\bar{a}\bar{b}} \sin(\Delta\delta -\Delta\phi) ~,\label{ATATbar}
\end{eqnarray}
with 
\begin{eqnarray}
{\cal I}_{ab}  & \propto & \frac{\int_R d\Phi \cdot f_{a } f_{ b} \cdot V \cdot  {\tt sign}(\tp)}{\int_R d\Phi \cdot f_{a } f_{ b} \cdot \left( W+ U \cdot \cos(\Delta\delta + \Delta\phi) \right)}  ~, \label{Iab}
\end{eqnarray}
where $d \Phi$ is the phase-space volume element, $R$ is the phase-space region of integration and 
$f_a,f_b$ are the PDF's of the incoming particles $a,b$; similarly, for the CC channel, ${\cal I}_{\bar{a}\bar{b}}$ is obtained by replacing  $f_a f_b \to f_{\bar a}f_{\bar b}$, $\tp \to \overline\tp$ 
and $\Delta\phi \to - \Delta\phi$.

As mentioned earlier, we see that $A_T\neq 0$ and/or $\bar{A}_T \neq 0$ can be observed even in the absence of CPV (i.e., when $\Delta\phi =0$), due to the presence of CP-even phases ($\Delta\delta \neq 0)$. 
Also, $|A_T| \neq |\bar{A}_T|$ is possible at the LHC, even with $\Delta\delta =0$, due the different PDF's of the incoming particles in the process and its CC channel, i.e., 
due to $f_a,f_b \neq f_{\bar a},f_{\bar b}$, when the initial state is not self conjugate.
This affects the CP-asymmetry $A_{CP}$ of \eqref{ACP1}, which is given by (using \eqref{ATATbar}):
\begin{eqnarray}
A_{CP}= \frac{{\cal I}_{ab} + {\cal I}_{\bar a \bar b}}2 \cos\Delta \delta \sin \Delta \phi + \frac{{\cal I}_{ab} - {\cal I}_{\bar a \bar b}}2 \sin \Delta \delta \cos \Delta \phi \cr~.
\label{ACP12}
\end{eqnarray}

Thus, when the initial state is self-conjuate and ${\cal I}_{ab} = {\cal I}_{\bar a \bar b}$ (i.e, the initial state and its CC state have the same PDF's),
then the asymmetry 
appears with the conventional CP-even and CP-odd phase factors, $A_{CP} \propto  \cos{\Delta\delta} \sin{\Delta\phi}$; in this case $A_{CP}$ vanishes  when the CP-odd phase vanishes. The second term in \eqref{ACP12}, which is $\propto {\cal I}_{\bar{a}\bar{b}} - {\cal I}_{ab}$, deals  with the case when the initial state is not self-conjugate 
and ${\cal I}_{ab} \neq {\cal I}_{\bar a \bar b}$, as is the case for the LHC or other future hadron colliders that are being envisioned (see also below).
This term is a new correction to the classic expression for tree-level CPV in scattering processes. It is a ``fake'' CP signal (being $\propto \cos\Delta\phi$) that can be generated in the presence of a CP-even phase. We note, though, that such a fake CP effect cannot be generated at tree-level in scattering processes at the LHC if there are no resonances involved (for situations involving resonances, see~\cite{Eilam:1991yv}), since then CP-even phases can only arise from FSI at higher loop orders, as opposed to the potentially large {\em tree-level} effects in $A_{CP}$, i.e., the 1st term in \eqref{ACP12}. It thus follows that, in the absence of resonances, if a large  CP asymmetry is measured, say of ${\cal O}(10\%)$, (as shown below), then besides the fact that it will be strong evidence for NP, it will also be a signal of  genuine CP-violating tree-level dynamics.

We  use an effective field theory (EFT) approach to describe the underlying NP responsible for CPV and demonstrate our strategy using the following scalar and tensor 4-Fermi operators \cite{EFT1,EFT2,EFT3,EFT4}:
\begin{eqnarray}
{\cal O}_S(prst) &=& 
(\bar l_p^j e_r) \epsilon_{jk} (\bar q_s^k u_t) 
\label{eq:Olequ1}~, \\
{\cal O}_T(prst) &=& 
(\bar l_p^j \sigma_{\mu\nu} e_r) \epsilon_{jk} (\bar q_s^k \sigma^{\mu\nu} u_t) \label{eq:Olequ3}~,
\end{eqnarray}
where $\ell$ and $q$ are left-handed SU(2) lepton and quark doublets, respectively; $e$ and $u$ are SU(2) singlet charged leptons and up-type quarks, respectively; and $p,r,s,t$ are flavor indices. These 4-Fermi interactions can be generated by tree-level exchanges of heavy scalars and tensors in the underlying heavy theory.  
Interesting examples are 
the scalar leptoquarks $S_1$ and $R_2$, which transform as $(3,1,-1/3)$ and $(3,2,7/6)$, respectively, under the $\textrm{SU(3)} \times \textrm{SU}(2) \times \textrm{U}(1)$ SM gauge group.\footnote{Note that the leptoquark $R_2$ is the only scalar leptoquark that does not induce proton decay~\cite{Dorsner:2016wpm}.} 
Indeed, these scalar leptoquarks can address the $R_{D^{(*)}}$ anomaly \cite{S1-S3-R2-1,S1-S3-R2-2,S1-S3-R2-3,S1-S3-R2-4,S1-S3-R2-5,S1-S3-R2-6,S1-S3-R2-7}, 
as well as the muon $g-2$ discrepancy~\cite{Fajfer:2021cxa,Aebischer:2021uvt} (see also~\cite{NeubertPRL,R2_1,R2_2,R2_3,R2_4,R2_5,R2_6,R2_7,R2_n,LQ-anom1} and for an alternative scenario with R-Parity violating Supersymmetry see~\cite{soniRPV1,soniRPV2,soniRPV3,Afik:2022vpm}). 

In particular, tree-level exchanges of $S_1$ and $R_2$ among the lepton-quark pairs induce the operators in \eqref{eq:Olequ1} and \eqref{eq:Olequ3},
where, in this case, the Wilson coefficients, $f_i$, of the operators in \eqref{eq:Olequ1} and \eqref{eq:Olequ3}, satisfy
\begin{eqnarray}
\vert f_T (prst) \vert = \frac{1}{4}\vert f_S(prst) \vert ~, \label{fTfS}
\end{eqnarray}
universally for any given set of flavor indices $prst$ in \eqref{eq:Olequ1} and \eqref{eq:Olequ3}, see~\cite{ourttllpaper}. We will use this relation as a benchmark scenario in the numerical calculations described below.

The scalar and tensor 4-Fermi operators in \eqref{eq:Olequ1} and \eqref{eq:Olequ3} (or equivalently, tree-level exchanges of the leptoquarks  $S_1$ and $R_2$) generate 
$t \bar{t} \ell^+ \ell^-$ as well as FC 
$t \bar{u}_i \ell^+ \ell^-$ (and the charge-conjugate $\bar{t} u_i \ell^+ \ell^-$) contact terms, where $\ell=e,\mu,\tau$ stands for any one of the SM charged leptons and $u_i=u,c$. 
The  $tt \ell \ell$ interaction modifies the process $pp \to t \bar t \ell^+ \ell^-$, as discussed in detail in~\cite{ourttllpaper}, and can thus also give rise to 
tree-level CPV in both the tri-lepton and four-leptons production channels of \eqref{3l_fs}-\eqref{4l_fs}.

\begin{figure}[H]
  \centering
\includegraphics[width=0.50\textwidth]{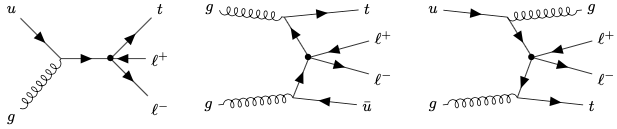}
\caption{Representative lowest order Feynman diagrams for $pp \to t \ell^+ \ell^-$ and $pp \to t \ell^+ \ell^- + j$ ($j$ is a light jet), via the $t u \ell^+ \ell^-$ 4-Fermi interaction (marked by a heavy dot).}
  \label{fig:Feynman}
\end{figure}

In the following, we focus
just on the FC $tu_i\ell\ell$ 
4-Fermi interactions, which can modify (see also~\cite{LFU-our,tull-our}) and generate CPV in the tri-lepton signals of \eqref{3l_fs} and \eqref{3l_fs_bar}, via the underlying single-top hard processes $u_i g \to t \ell^+ \ell^- $ and the CC channel (see Fig.~\ref{fig:Feynman}), 
followed by the $t$ and $\bar{t}$ decays $t \to b \ell^{\prime +} \nu_\ell$ and $\bar{t} \to \bar{b} \ell^{\prime -} \bar{\nu}_\ell$.

As discussed below, the contribution of the FC $tu_i\ell\ell$ effective operators to the tri-lepton signal does not interfere with the SM diagrams, so that the CPV in this case is a pure NP effect; it arises from the imaginary part of the interference between the scalar and the tensor operators, if at least one of the corresponding Wilson coefficients 
is complex.\footnote{Equivalently, assuming that the underlying scattering processes/diagrams are generated by the tree-level exchanges of the two leptoquarks $S_1$ and $R_2$, the CP-violating effect arises from the interference between them (i.e., the amplitudes $M_1$ and $M_2$ in \eqref{M} are generated by $S_1$ and $R_2$, respectively) and if their couplings to a $t \ell$ and $u\ell$ (and/or $c\ell$) pairs carry a (different) CP-violating phase. Interesting examples of CPV leptons and quarks Electric Dipole Moments (EDM's), which are generated from $S_1$ and $R_2$ complex couplings can be found in \cite{Dekens:2018bci,S1R2_CPV}. }

In particular, the numerator of $A_{CP}$ (and of $A_T$ and $\bar{A}_T$) is proportional to the CP-violating part of the cross-section for
$u_i g \to t \ell^+ \ell^- \to \ell^{\prime +} \ell^+ \ell^- +X$ (hereafter we suppress the flavor indices of the operators in \eqref{eq:Olequ1} and \eqref{eq:Olequ3}):\footnote{Note that for a given lepton flavor, anyone of the FC 4-Fermi operators has two coupling products contributing to $d \hat\sigma(CPV)$ in \eqref{sigCPV}, which correspond to different quark indices.  For example, in the case of the $tc \mu \mu$ interaction, we denote by ${\rm Im}(f_S f_T^\star)$ any one of the products 
${\rm Im} \left(f_S(2232) \cdot f_T^\star(2232) \right)$ or ${\rm Im} \left(f_S^\star(2223) \cdot f_T(2223)\right)$; only one of the two will be "turned on" henceforward.}
\begin{eqnarray}
   d\hat\sigma(CPV) \propto \epsilon \left( p_{u_i},p_{\ell^{\prime +}},p_{\ell^+},p_{\ell^-} \right) \cdot {\rm Im} 
   \left( f_S f_T^\star \right)~, \label{sigCPV}
\end{eqnarray}
and similarly for the CC channel $\bar{u}_i g \to \bar{t} \ell^- \ell^+ \to \ell^{\prime -} \ell^- \ell^+ + \bar{X}$ by replacing $\epsilon \left( p_{u_i},p_{\ell^{\prime +}},p_{\ell^+},p_{\ell^-} \right)$ with $\epsilon \left( p_{\bar{u}_i},p_{\ell^{\prime -}},p_{\ell^-},p_{\ell^+} \right)$, 
where $\epsilon \left( p_1,p_1,p_3,p_4 \right) = \epsilon_{\alpha \beta \gamma \delta} p_1^\alpha p_2^\beta p_3^\gamma p_4^\delta$ and $\epsilon_{\alpha \beta \gamma \delta}$ is the Levi-Civita tensor. 
In contrast to the numerators, the NP contributions to the denominators of our CP-asymmetries are proportional to the CP-conserving terms $\propto |f_S|^2, |f_T|^2, {\tt Re}(f_S \cdot f_T^\star)$, where the dominating term is the pure tensor contribution $|f_T|^2$. The SM tri-lepton production processes will also contribute to the total number of tri-lepton events which enter the denominators of $A_{CP}$ and $T_N,\bar{T}_N$; the dominating SM tri-lepton process is $pp \to WZ +X$.\footnote{Other irreducible SM background to the inclusive tri-lepton signals are $pp \to t \bar t V$ with $V=W,Z$ and  $pp \to t \bar t t \bar t$. These, however, are more than an order of magnitude smaller than the $WZ$ one in the inclusive channel. Note, though, that the $t \bar t V$ and $t \bar t t \bar t$ backgrounds may become important if specific selections are used, e.g., $b$-jet tagging.} 

To assess the feasibility of CP asymmetry measurements in multi-leptons final states at the LHC, we perform a simulation on the tri-lepton signal processes described above, together with the relevant SM background processes, which do not include detector effects other than those modeled by simple threshold and acceptance requirements.
Although more elaborated analysis approaches might also be useful, for simplicity, we follow an approach that is completely generic and provides a model-independent test of CPV in multi-lepton final states, which would be designed to be sensitive to any type of underlying CP-violating NP involving charged-leptons. We therefore define the asymmetries for the inclusive multi-lepton signals, with no further event selections on the types or kinematic properties of the other objects in the final state, i.e., $X_i$ in \eqref{3l_fs}-\eqref{4l_fs}.
Indeed, in general it is possible to use additional useful selections , e.g., in our case a selection of one $b$-jet (see~\cite{bsll-our,bbll-our,tull-our,ourttllpaper,ATLAS:2021mla}) will essentially eliminate the dominating $pp \to W^\pm Z +X \to \ell^{\prime \pm} \ell^+ \ell^- +X$ SM contribution to the denominators of our asymmetries. 
Nonetheless, we use only a selection on the minimum invariant mass of the di-leptons involved, 
$m_{\tt min}(\ell^+ \ell^-)$, which allows us to suppress the SM background without loss of generality.
The input for the numerical calculations is further described in Appendix \ref{apA}.

Furthermore, 
for the NP contribution we  study the dependence on the NP scale up to $\Lambda \sim $ few TeV; the typical  bounds on the natural scale of the operators under investigation, in \eqref{eq:Olequ1} and \eqref{eq:Olequ3}, are $\Lambda \gsim {\cal O}(1)$~TeV, see~\cite{tull-our}. Guided by the relation between the scalar and tensor couplings in \eqref{fTfS}, we set $\vert f_S \vert = 1$, $\vert f_T \vert = 0.25$ with a maximal CP-odd phase for the $t u \ell \ell$ and $t c \ell \ell$ operators, so that:
\begin{equation}
    {\rm Im} \left( f_S \cdot f_T^\star \right) = 0.25 ~. \label{CPVvalue}
\end{equation}

\begin{figure}[]
  \centering
\includegraphics[width=0.50\textwidth]{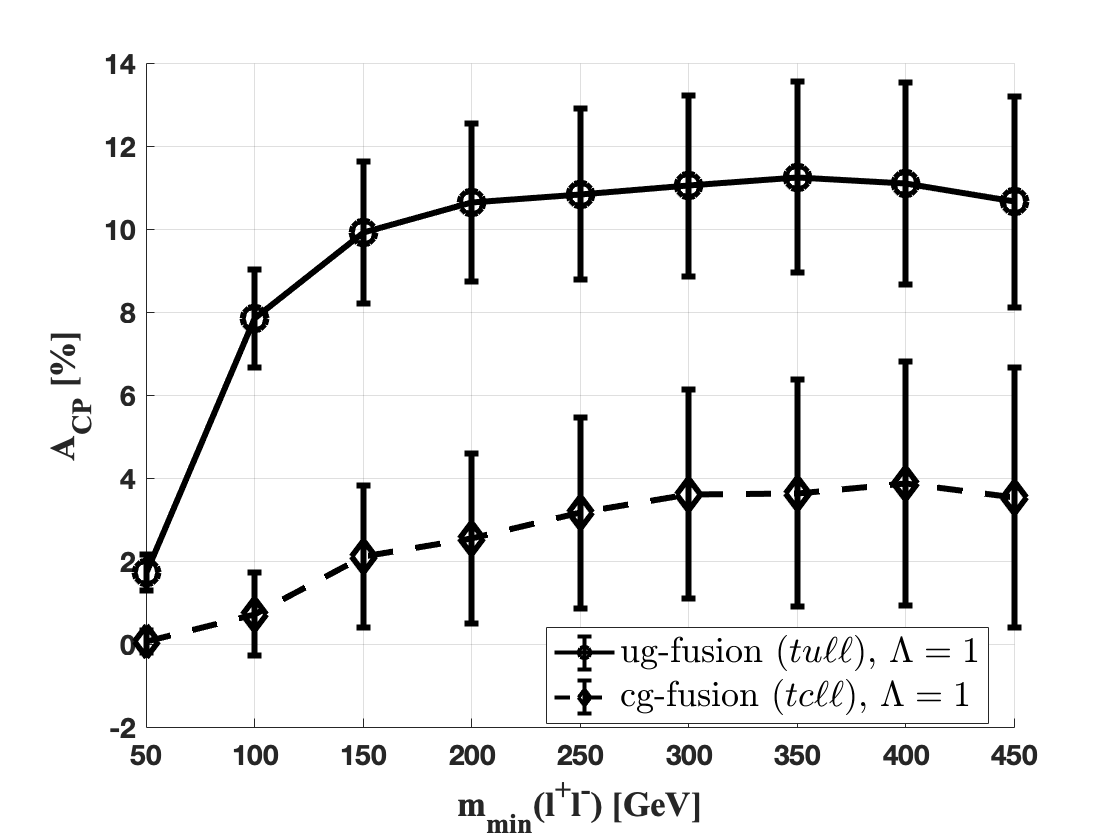}
\caption{$A_{CP}$ as a function of $m_{min}(\ell^+ \ell^-)$, for $\Lambda=1$ TeV, ${\rm Im} \left(f_S f_T^\star \right) =0.25$ and including the SM background. The dependence of the asymmetry on $\Lambda$ is given in Appendix \ref{apB}. The error bars represent the expected statistical uncertainty with an integrated luminosity of $1000(3000)$ fb$^{-1}$ for the ug-fusion(cg-fusion) case.
}
  \label{fig:ACP}
\end{figure}

Our results  are summarized in Fig.~\ref{fig:ACP} and Table~\ref{tab:data2}.
In Fig.~\ref{fig:ACP} we show the dependence of $A_{CP}$ on $m_{\tt min}(\ell^+ \ell^-)$ and in
Table~\ref{tab:data2} we give the resulting CP-violating and $T_N$-odd asymmetries for $m_{\tt min}(\ell^+ \ell^-)=400$~GeV. 
The expected inclusive tri-lepton cross-sections for the NP and the dominant SM background, after the event selection criteria have been applied, are given in Appendix \ref{apB}:
for $m_{\tt min}(\ell^+ \ell^-) = 400$~GeV and an integrated luminosity of $1000$ fb$^{-1}$, we expect an ${\cal O}(100)$ $\ell^{\prime \pm} \ell^+ \ell^-$ from the SM $pp \to ZW^\pm$ background, whereas the new $tu \ell \ell$ and $tc \ell \ell$ 4-Fermi operators  yield $ \sim 10^4$ and $\sim 500$ $\ell^{\prime \pm} \ell^+ \ell^-$ events, respectively, if $\Lambda \sim 1$~TeV.

We see that the CP-asymmetry increases with the invariant mass cut on the same-flavor di-leptons, $m_{\tt min}(\ell^+ \ell^-)$. This is due to the decrease of the SM contribution with $m_{\tt min}(\ell^+ \ell^-)$ in the denominators of the asymmetries.
Also, the asymmetry is larger in the $ug$-fusion case, since the SM background in this case is considerably smaller w.r.t. the signal in this case (see the Appendix and discussion above).\footnote{The uncertainty (numerical) in the reported asymmetries is of ${\cal O}(0.1\%)$. This is estimated by "turning off" the CP-violating NP contribution and calculating $A_T, \bar{A}_T$ and $A_{CP}$ within the SM, where it is expected to vanish.} 
On the other hand, the asymmetries $A_T,\bar{A}_T$ and $A_{CP}$ decreases with $\Lambda$, as expected. For example, in the $tu \ell \ell$ 4-Fermi case, the CP-asymmetry drops from $A_{CP} \sim 11\%$ if $\Lambda =1$~TeV to $A_{CP} \sim 8\%$ if $\Lambda =2$~TeV (see Table~\ref{tab:data2}). A plot of $A_{CP}(\Lambda)$
is given in Appendix \ref{apB}. Note also that $|A_T| >> |\bar{A}_T|$ in the $ug$-fusion case due to the difference between the incoming $ug$ and $\bar{u}g$ PDF's, see \eqref{ATATbar}.
\begin{table}[htb]
\caption{The expected $T_N$-odd and CP asymmetries in tri-lepton events, $pp \to \ell^{\prime \pm} \ell^+ \ell^- +X$, via the $ug$-fusion and $cg$-fusion production channels (and the CC ones) at the LHC, for $m_{min}(\ell^+ \ell^-) = 400$~GeV. Values are given for $\Lambda=1(2)$~TeV, ${\rm Im} \left(f_S f_T^\star \right) =0.25$ and the SM background from $pp \to ZW^{\pm} +X$, as explained in the text.
\label{tab:data2}}
\begin{tabular}{c|c|c}
 & $ug$-fusion: $\Lambda=1(2)$~TeV & $cg$-fusion: $\Lambda=1(2)$~TeV 
\tabularnewline
\hline \hline
\
$A_{CP}$ & $11.1$\% (7.9)\% & $3.9$\% (0.7)\% \tabularnewline
\hline 
\
$A_{T}$ & $16.4$\% (13.5)\% & $3.1$\% (0.5)\% \tabularnewline
\hline 
\
$\bar{A}_T$ & $-5.8$\% (-2.3)\% & $-4.7 $\% (-1.0)\% \tabularnewline
\hline \hline
\end{tabular}
\end{table}

\newpage 

Finally, it is possible to further refine this approach by defining the axis-dependent TP CP-asymmetries $\tp^i = p_{a}^i \cdot \left( \Vec{p}_b \times \Vec{p}_c \right)^i$, where $i=x,y,z$. As shown in Appendix \ref{apC}, the $\tp^{x,y,z}$ can be useful for a deeper understanding of the origin of the underlying CP-violating NP; in the case of the 4-Fermi effective interactions studied here, they allow us to distinguish between the $tu \ell \ell$ and the $tc \ell \ell$ CP-violating dynamics.

To summarize, we have investigated the possible detection of tree-level CPV in scattering processes at the LHC and introduced a modification to the standard formula for such CP-violating effects, which is relevant when the initial state in not self-conjugate. We focused specifically on multi-leptons signals and their sensitivity to new TeV-scale sources of CPV. In particular,  
we have constructed CP-violating triple-product correlations out of the momenta of the charged leptons in multi-lepton events, which can be used as model-independent tests of tree-level (and therefore large) CPV from any source of underlying CP-violating physics. We have calculated the expected CP-asymmetry in tri-lepton events at the LHC from new TeV-scale FC $tu \ell \ell$ and $tc \ell \ell$ 4-Fermi interactions, which can be viewed as an EFT parameterization of tree-level TeV-scale leptoquark exchanges in these channels. We showed that an ${\cal O}(10\%)$ CP-asymmetry is naturally expected in this case, if the EFT operators carry a CP-odd phase and the NP scale is of ${\cal O}(TeV)$. 

The measurement of such ${\cal O}(10\%)$ CP-asymmetry in multi-lepton events is challenging, but 
if observed, it should stand out as an unambiguous signal of NP that may shed light on the fundamental issue of BAU. 
We believe that it is quite feasible provided the experimental uncertainties can be kept
at the level of ${\cal O}(1\%)$ (see~\cite{CMS:2022voq})
bearing in mind that such CP-violating effects in the SM are un-observably small in multi-lepton events. Indeed, we estimate  the statistical uncertainty in measuring the CP-asymmetry, based on the expected number of tri-lepton events in our NP scenario (see Appendix)  to be 
 $\sim 1\% - 2\%$ with an integrated luminosity of  ${\cal L} \sim 1000(3000)$ fb$^{-1}$ in the $tu\mu\mu(tc\mu\mu)$ NP cases   
(see Fig.~\ref{fig:ACP}). 

\acknowledgments
We thank Ilaria Brivio for useful help with the MC simulations.
The work of AS  was supported in part by the U.S. DOE contract \#DE-SC0012704. YA is supported by the National Science Foundation under Grant No. PHY-2013010.

\clearpage

\appendix

\section{Numerical calculations \label{apA}}

All event samples (NP signal and SM background) were generated using {\sc MadGraph5\_aMC@NLO}~\cite{madgraph5} at LO parton-level and with the SMEFTsim model of~\cite{SMEFTsim1,SMEFTsim2} for the EFT framework. 
The 5-flavor scheme was used to generate all samples, with the ${\tt NNPDF30\_lo\_as\_0130}$ PDF set~\cite{Ball:2014uwa} and 
the default {\sc MadGraph5\_aMC@NLO} LO dynamical scale.

Both the NP and the SM tri-lepton production cross-sections were calculated with an additional jet. In particular, for the NP: 
$pp \to t(\bar t) \ell^+ \ell^-$ and $pp \to t(\bar t) \ell^+ \ell^- +j$  
followed by the top(anti-top) decay $t(\bar t) \to b \ell^{\prime +} \nu_{\ell^\prime} (\bar b \ell^{\prime -}\bar\nu_{\ell^\prime})$, while for the SM: $pp \to Z W^\pm$ and $pp \to Z W^\pm + j$ followed by 
$Z \to \ell^+ \ell^-$ and 
$W^\pm \to \ell^{\prime \pm} \nu_{\ell^{\prime}}$. 

Leptons were required to have transverse momentum of $p _ {\mathrm{T} } >10$~GeV and pseudo-rapidity $\left|\eta\right|<2.5$, while for jets we used $p _ {\mathrm{T} } >20$~GeV, $\left|\eta\right|<5.0$ and an angular separation of $\Delta R=0.4$.

In Table~\ref{tab:data1} we list the estimated cross-sections for the NP with $\Lambda=1$~TeV (note that the NP cross-section scales as $\Lambda^{-4}$) and the SM contributions to the inclusive 
$pp \to \ell^{\prime \pm} \ell^+ \ell^- + X $ processes for $m_{min}(\ell^+ \ell^-)=200,~300$ and $400$~GeV, where $m_{min}(\ell^+ \ell^-)$ is the lower cut on the invariant mass of the same-flavor di-leptons. In particular, the $m_{min}(\ell^+ \ell^-)$-dependent cross-sections are defined as:
\begin{eqnarray}
\sigma_{m_{min}(\ell^+ \ell^-)} \equiv 
\int_{m(\ell \ell) \geq m_{min}(\ell^+ \ell^-)} d m (\ell \ell) \frac{d\sigma}{dm(\ell \ell)} ~. \label{CCSX}
\end{eqnarray}

We note that the simulations were made without parton showering and jet matching, which has no effect on our CP-asymmetry (we confirmed that the calculated  CP-asymmetry with and without the extra jet in the tri-lepton final state is the same within the numerical error). Also,  we did not perform any detector simulation which is beyond the scope of this work and is left for a dedicated analysis. Thus, the cross-sections reported in Table~\ref{tab:data1} should be viewed as an estimate; a more realistic calculation of the expected total cross-sections for this type of NP and SM background can be found in~\cite{tull-our}.

\begin{table}[]
\caption{The estimated cross-sections in [fb], for the NP tri-lepton signals and the SM tri-lepton background.
Values are given for the NP parameters ${\rm Im} \left(f_S f_T^\star \right) =0.25$, $\Lambda=1$~TeV and for three values of $m_{min}(\ell^+ \ell^-)$ as indicated. Also, all acceptance cuts (e.g., $p_T$ and $\eta$ of the leptons) have been applied, see also description in the text.\label{tab:data1}}
\begin{tabular}{c|c|c|c}
$m_{min}(\ell^+ \ell^-) [GeV] \Rightarrow$ & $200$ & $300$ & $400$  
\tabularnewline
\hline \hline
\
$\sigma_{NP}(pp_{ug} \to \ell^{\prime -} \ell^+ \ell^- +X) $ & 12.43 & 11.65 & 10.84 \tabularnewline
\hline 
\
$\sigma_{NP}(pp_{\bar u g} \to \ell^{\prime +} \ell^- \ell^+ +X) $ & 0.98 & 0.87 & 0.76
\tabularnewline
\hline 
\
$\sigma_{NP}(pp_{cg} \to \ell^{\prime -} \ell^+ \ell^- +X) $ & 0.37 & 0.32 & 0.27 \tabularnewline
\hline 
\
$\sigma_{NP}(pp_{\bar c g} \to \ell^{\prime +} \ell^- \ell^+ +X) $ & 0.37 & 0.32 & 0.27
\tabularnewline
\hline 
\
$\sigma_{SM}(pp \to \ell^{\prime -} \ell^+ \ell^- +X) $ & 0.33 & 0.11 & 0.05
\tabularnewline
\hline 
\
$\sigma_{SM}(pp \to \ell^{\prime +} \ell^- \ell^+ +X) $ & 0.56 & 0.21 & 0.10
\tabularnewline
\hline \hline

\end{tabular}
\end{table}

\section{Dependence of the CP-asymmetry on the NP scale \label{apB}}

\begin{figure}[]
  \centering
\includegraphics[width=0.50\textwidth]{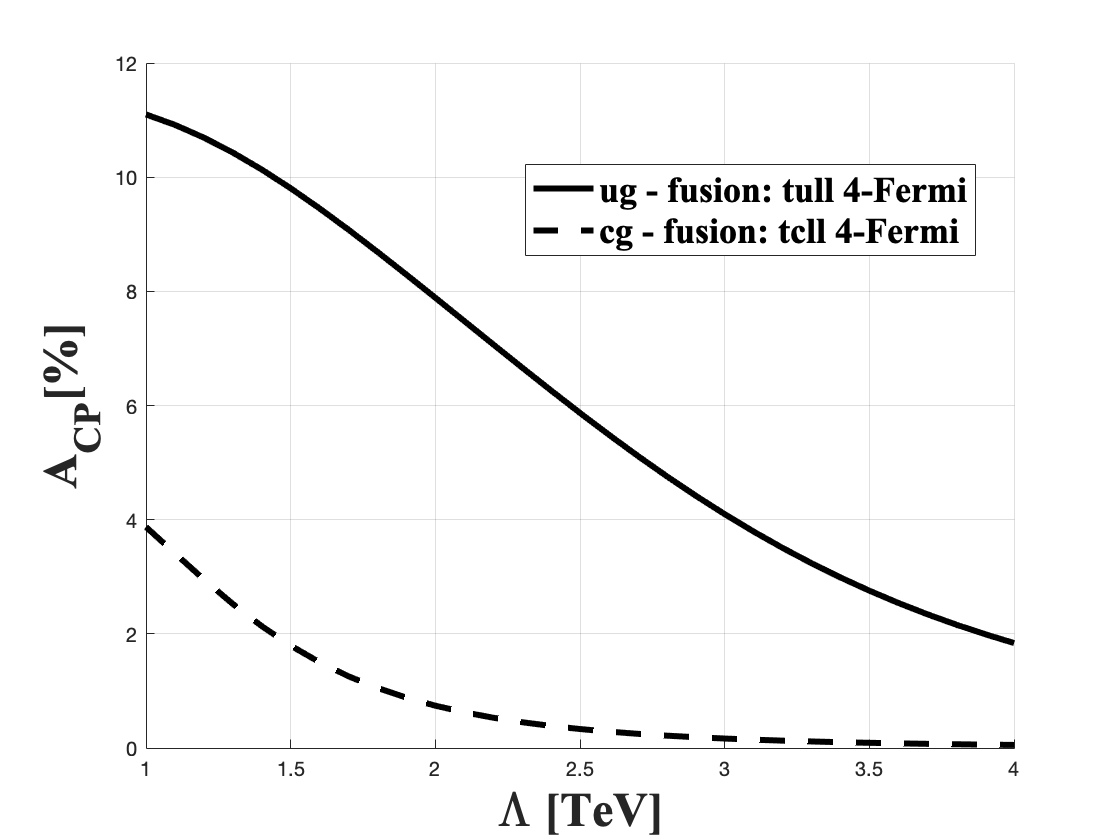}
\caption{The expected CP-asymmetry $A_{CP}$, as a function of the NP scale $\Lambda$, for
$m_{min}(\ell^+ \ell^-)=400$~GeV and ${\rm Im} \left(f_S f_T^\star \right) =0.25$. 
Results are shown for the cases of NP from $ug$ and $cg$-fusion, which arise from 
the $tu\ell\ell$ and $tc \ell\ell$ 4-Fermi operators, respectively.
The SM background is calculated from $pp \to ZW^{\pm} +X$.}
  \label{fig:ACP_Lambda}
\end{figure}

In Fig.\ref{fig:ACP_Lambda} we show the dependence of $A_{CP}$ on the NP scale $\Lambda$, with 
parameters as indicated in the caption of the figure. We see that the asymmetry in the ug-fusion case falls rather slowly in the range $\Lambda \sim 1-4$ TeV, whereas in the cg-fusion case it drops steeply in this range, approaching $1/\Lambda^4$ where the SM contribution to the inclusive tri-lepton background dominates.

\section{Axis-dependent CP-violating triple product observables \label{apC}}

The triple products considered in the paper:\footnote{The triple products are defined in the laboratory frame and we expect that systematic uncertainties in the reconstruction of the momenta involved will be smaller then e.g., the case where the momenta are defined in a rest frame of some particle(s). Also, the kinematical cuts on the leptons involved should be CP-symmetric, e.g., same $p_T$ cuts should be applied to all leptons.}
\begin{equation}
    \tp = \Vec{p}_a \cdot \left( \Vec{p}_b \times \Vec{p}_c \right) \label{TPbasic1} ~,
\end{equation}
can be divided into three axis-sensitive triple-products:
\begin{equation}
    \tp^i = p_{a}^i \cdot \left( \Vec{p}_b \times \Vec{p}_c \right)^i \label{TPbasic2} ~,
\end{equation}
where $i=x,y,z$ denotes the 
$x,y,z$ components of the momenta, e.g., $p_{a}^z$ and 
$\left( \Vec{p}_b \times \Vec{p}_c \right)^z$ are the $z$-components of the momenta $\Vec{p}_{a}$ and 
$\left( \Vec{p}_b \times \Vec{p}_c \right)$, respectively.
Note that only three out of the four $\tp$ and $\tp^{x,y,z}$ in \eqref{TPbasic1} and \eqref{TPbasic2} are independent, since $\tp = \sum_{i=x,y,z} \tp^i$. Furthermore, the axis-sensitive $\tp^{x,y,z}$ transform under $P$,$C$,$CP$ and $T_N$ the same as $\tp$, so that all the discussion and formulae for $\tp$ in the paper applies also to $\tp^{x,y,z}$. In particular, the axis-dependent CP-asymmetries can be similarly defined as: 
\begin{eqnarray}
    A_{CP}^{x,y,z} = \frac{1}{2} \left(A_T^{x,y,z} - \bar{A}_T^{x,y,z} \right) \label{ACPX} ~,
\end{eqnarray}
where $A_T^{x,y,z}$ and $\bar{A}_T^{x,y,z}$ are the axis-dependent $T_N$-odd asymmetries.
\begin{table}[]
\caption{The expected $T_N$-odd and CP asymmetries $A_{T}$, $\bar{A}_T$, $A_{CP}$ and the corresponding axis-dependent asymmetries $A_{T}^i$, $\bar{A}_T^i$, $A_{CP}^i$ ($i=x,y,z$), for the tri-lepton events $pp \to \ell^{\prime \pm} \ell^+ \ell^- +X$ at the LHC with $m_{min}(\ell^+ \ell^-) = 400$~GeV. Results are given for both the $ug$-fusion and $cg$-fusion production channels (and the CC ones). Numbers are presented for $\Lambda=1$~TeV, ${\rm Im} \left(f_S f_T^\star \right) =0.25$ and the dominant SM background from $pp \to ZW^{\pm} +X$ is included. The cases where an asymmetry is $\lsim 0.5\%$ is marked by an X.  \label{tab:data3}}
\begin{tabular}{cc|c|c|c}
& $A_{CP}$ & $~A_{CP}^x$ & $A_{CP}^y$ & $A_{CP}^z$
\tabularnewline
\hline \hline
\
$ug$-fusion: & 11.1\% & 8.1\%, & 8.1\% & X \tabularnewline
\hline 
\
$cg$-fusion: & 3.9\% & X & X & 5.6\% \tabularnewline
\hline \hline
\tabularnewline
& $A_{T}$ & $~A_{T}^x$ & $A_{T}^y$ & $A_{T}^z$
\tabularnewline
\hline \hline
\
$ug$-fusion: & 16.4\% & 11.3\%, & 10.7\% & 3.8\% \tabularnewline
\hline 
\
$cg$-fusion: & 3.1\% & 5.0 & X & X \tabularnewline
\hline \hline
\tabularnewline
& $\bar{A}_T$ & $\bar{A}_T^x$ & $\bar{A}_T^y$ & $\bar{A}_T^z$
\tabularnewline
\hline \hline
\
$ug$-fusion: & -5.8\% & -5.0\% & -5.6\% & 3.1\% \tabularnewline
\hline 
\
$cg$-fusion: & -4.7\% & -6.3\% & X & X  \tabularnewline
\hline \hline
\end{tabular}
\end{table}

In Table~\ref{tab:data3} we show a sample of our results for all $T_N$-odd and CP-asymmetries 
including the axis dependent ones, for the 
tri-lepton events $pp \to \ell^{\prime \pm} \ell^+ \ell^- +X$ at the LHC, which are considered in this paper. 
The asymmetries are calculated for both the $ug$-fusion and $cg$-fusion production channels (and the CC ones), with $m_{min}(\ell^+ \ell^-) = 400$~GeV , $\Lambda=1$~TeV and ${\rm Im} \left(f_S f_T^\star \right) =0.25$, and the dominant SM background from $pp \to ZW^{\pm} +X$ is considered.
We see that a measurement of the axis-dependent asymmetries can be used to distinguish between the $tu \ell \ell$ and the $tc \ell \ell$ CP-violating dynamics. In particular, in the $tu \ell \ell$ case we obtain $A_{CP}^z \to 0$ and $A_{CP}^{x,y} \sim 8\%$, while for the $tc \ell \ell$ operator we find
$A_{CP}^z \sim 5.5\%$ and $A_{CP}^{x,y} \to 0$. 
Note also that the axis-dependent asymmetries may yield a larger effect, e.g., in the $cg$-fusion case we find that $A_{CP}^z > A_{CP}$. 

\section{Differential distributions: signal vs. background \label{apD}}

\begin{figure*}[]
\centering
\includegraphics[width=0.45\textwidth]{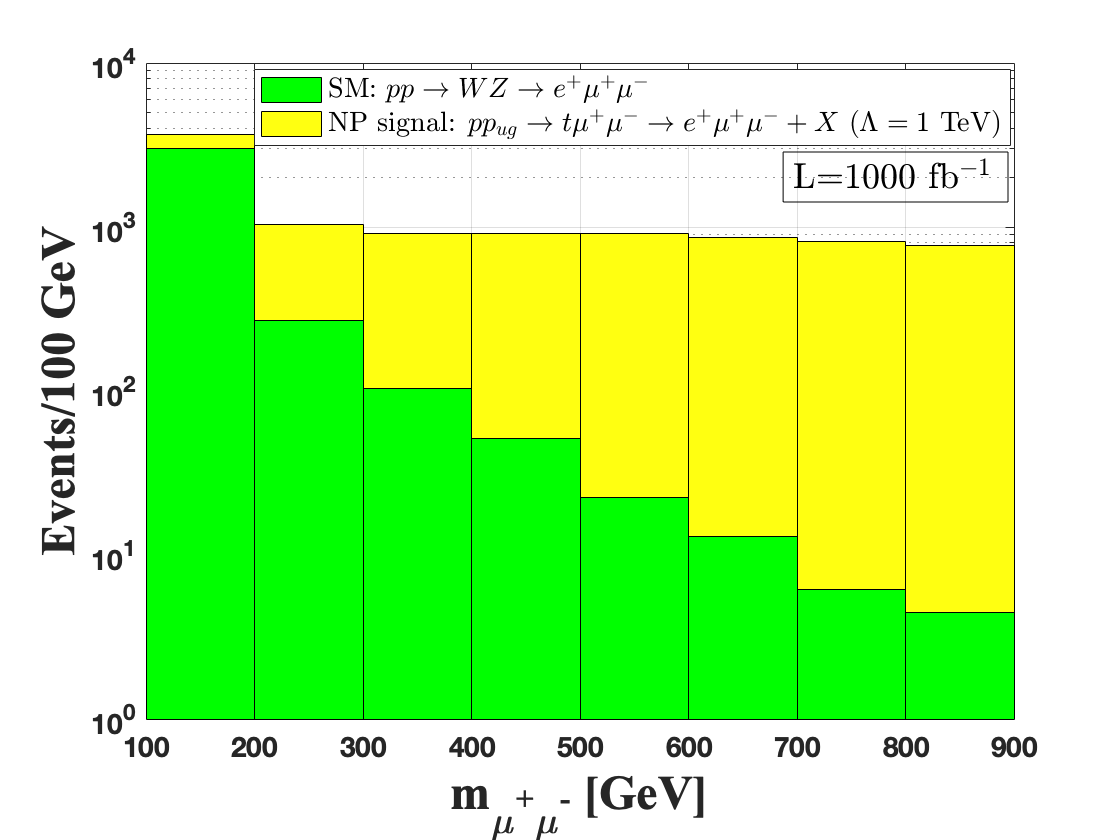}
\includegraphics[width=0.45\textwidth]{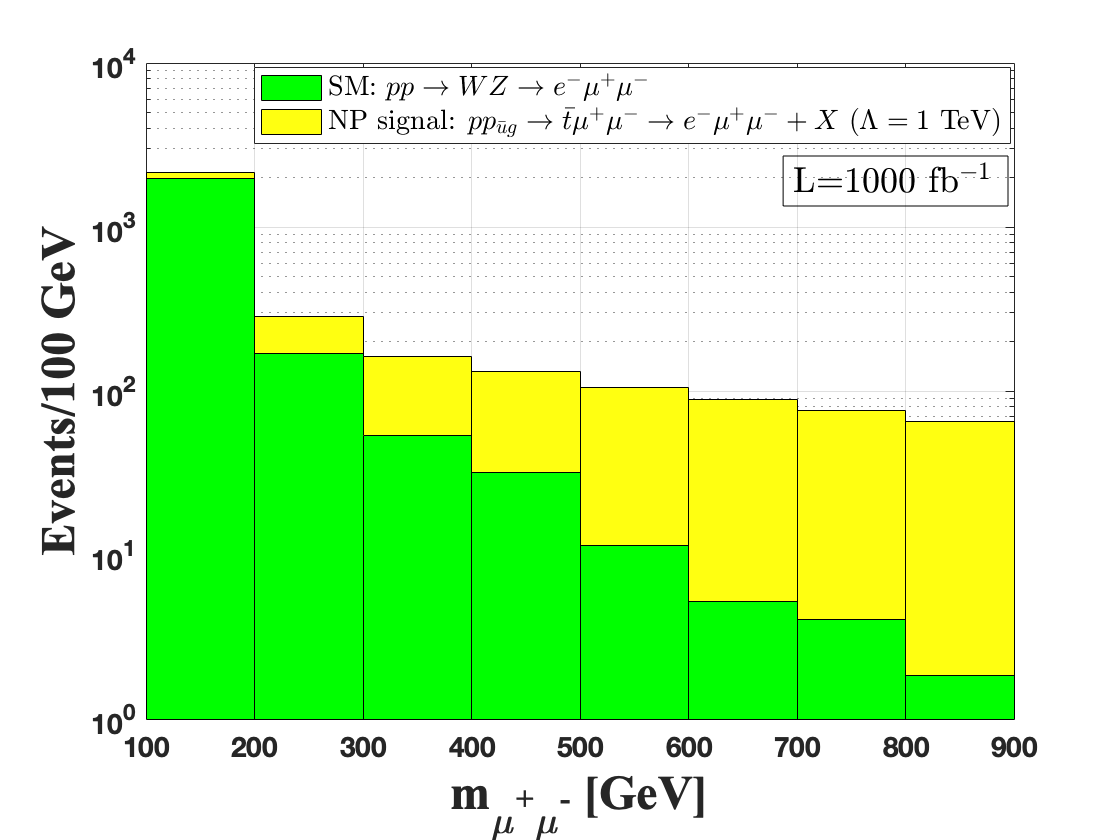}
\includegraphics[width=0.45\textwidth]{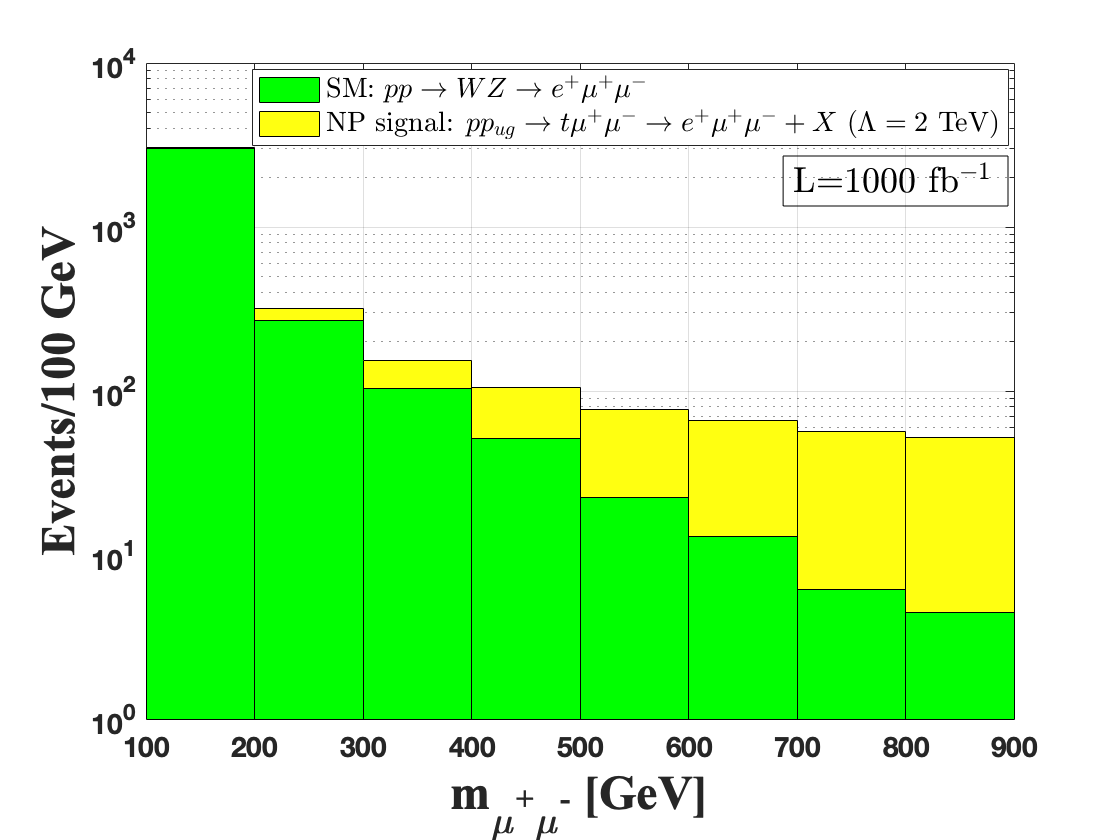}
\includegraphics[width=0.45\textwidth]{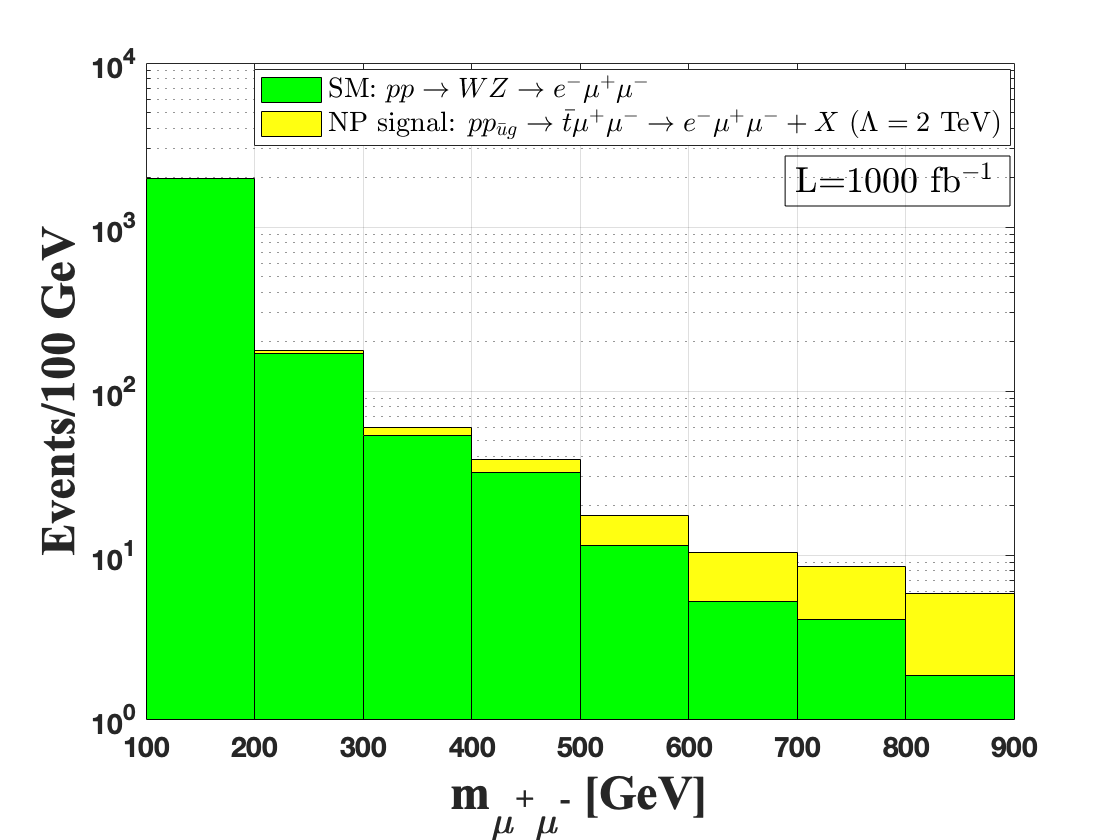}
\caption{Di-muon invariant mass distribution (stacked) for the tri-lepton $e^+ \mu^+ \mu^-$ (left figures) and $e^- \mu^+ \mu^-$ (right figures) signals, from $ug$-fusion and $\bar u g$-fusion NP processes, respectively, and the corresponding SM backgrounds. The distributions are shown per integrated luminosity of ${\cal L}=1000$~fb$^{-1}$, for $\Lambda=1$ TeV (upper figures) and $\Lambda=2$ TeV (lower figures).}
\centering
\label{fig:mlld_ist_Lam12}
\end{figure*}

\begin{figure*}[]
\centering
\includegraphics[width=0.45\textwidth]{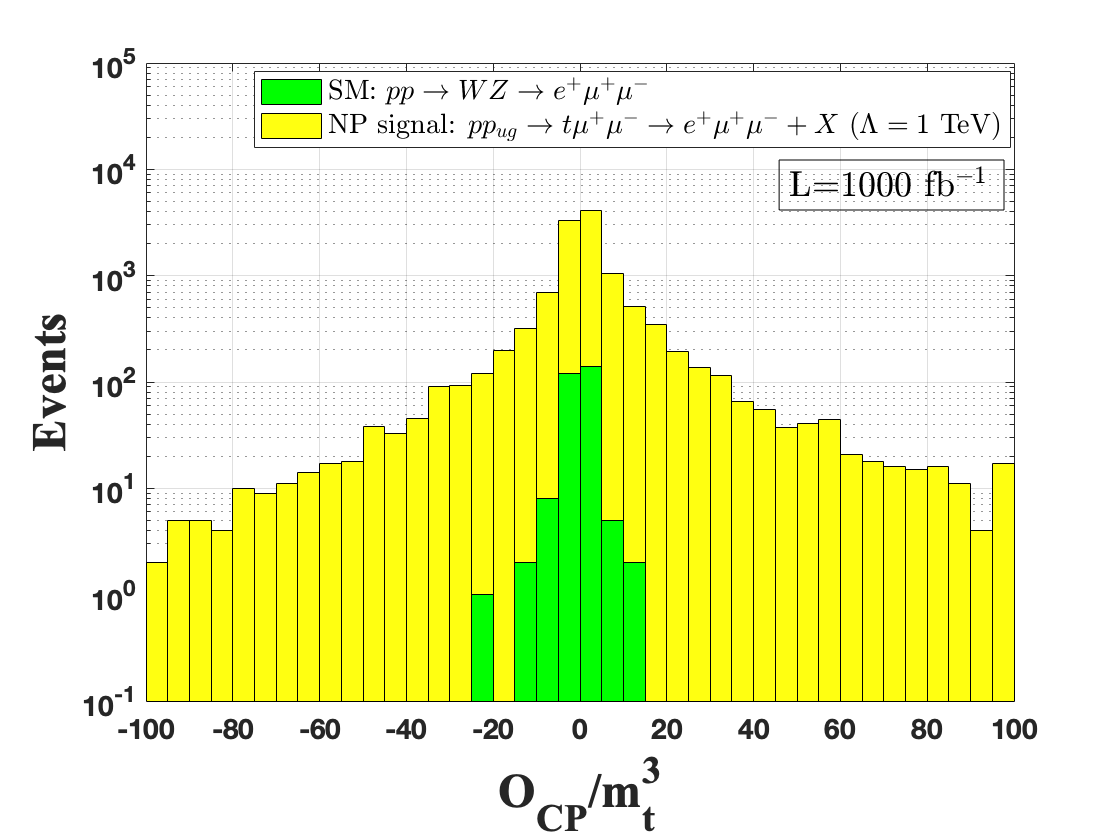}
\includegraphics[width=0.45\textwidth]{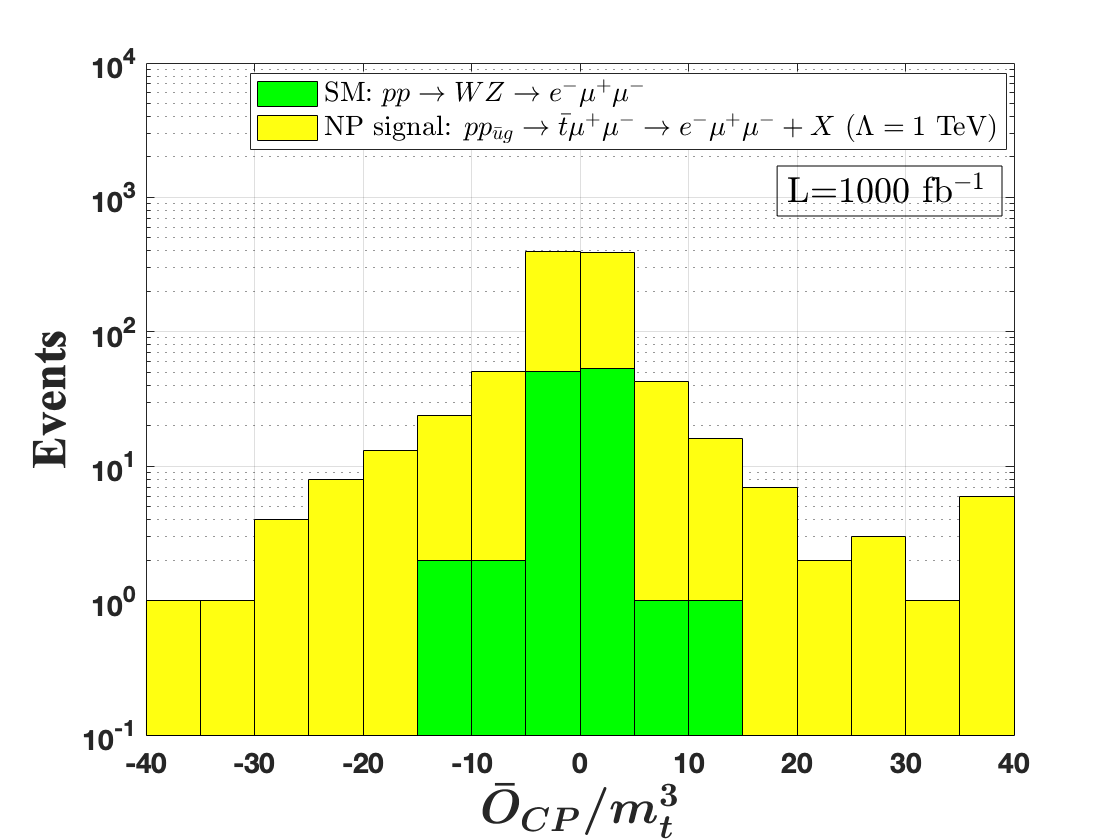}
\includegraphics[width=0.45\textwidth]{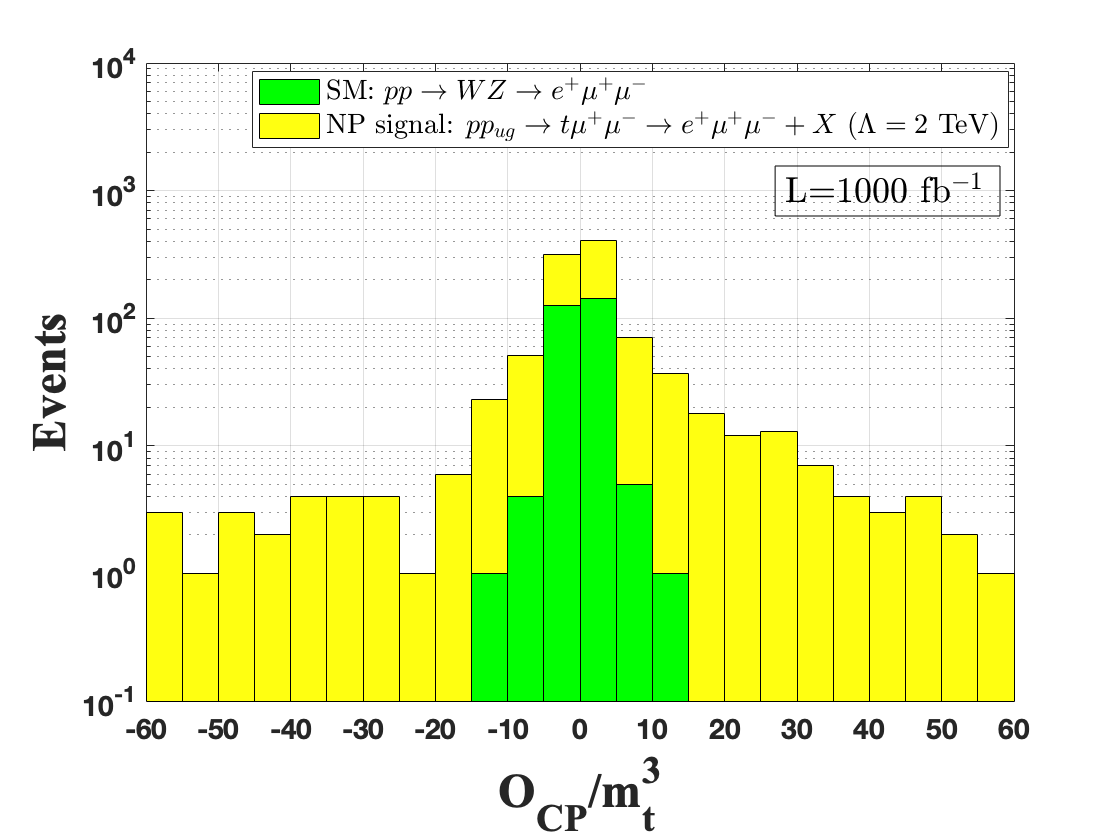}
\includegraphics[width=0.45\textwidth]{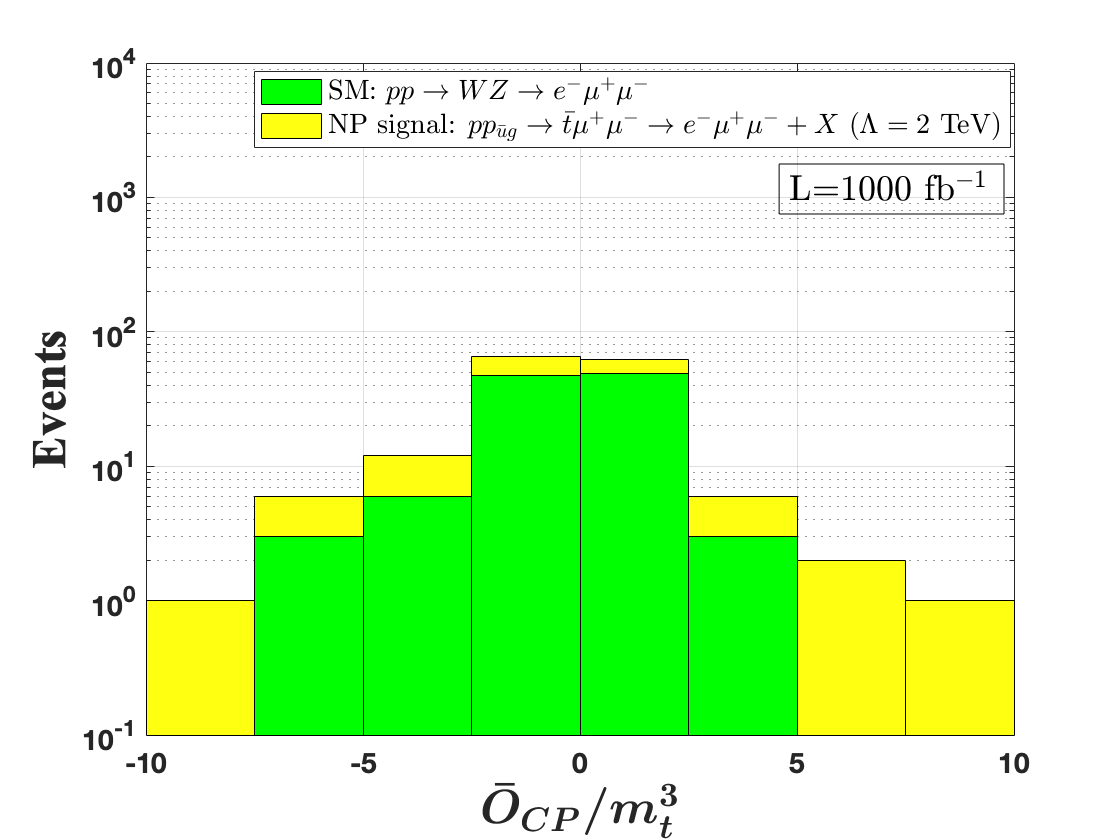}
\caption{Differential distribution (stacked) of the triple products ${\cal O}_{CP}$ (left figures) and $\bar{{\cal O}}_{CP}$ (right figures) in the tri-lepton NP signals and corresponding backgrounds. The NP is from $ug \to t \mu^+ \mu^- \to e^+ \mu^+ \mu^-$ (left figures) and $\bar{u} g \to \bar{t} \mu^+ \mu^- \to e^- \mu^+ \mu^-$ (right figures) with $\Lambda=1$ TeV (upper figures) and $\Lambda=2$ TeV (lower figures). 
The distributions for both signal and background are calculated with the cut on the di-muon invariant mass of 
$m_{min}(\ell^+ \ell^-)=300$~GeV and per integrated luminosity of ${\cal L}=1000$~fb$^{-1}$. See also text.}
\centering
\label{fig:TPd_ist_Lam12}
\end{figure*}

In Figs.~\ref{fig:mlld_ist_Lam12} and \ref{fig:TPd_ist_Lam12},
we plot the di-muon invariant mass and the triple-product differential distributions, respectively, for  an integrated luminosity of ${\cal L}=1000$~fb$^{-1}$. 
We show these distributions for the $ug$-fusion and the CC $\bar{u}g$-fusion NP cases and the corresponding SM backgrounds, assuming a NP scale of $\Lambda=1$ TeV and/or $\Lambda=2$ TeV. Note that the NP signals scale as $\Lambda^{-4}$ and 
are calculated with our benchmark 
value for the CPV coupling ${\rm Im} \left(f_S f_T^\star \right) =0.25$. Also, the triple-product distributions in Fig.~\ref{fig:TPd_ist_Lam12} are calculated with 
$m_{min}(\ell^+ \ell^-)=300$~GeV. 

%
%\begin{figure}[]
%\centering
%\includegraphics[width=0.4\textwidth]{mll_dist_ug_Lam2.png}
%\includegraphics[width=0.4\textwidth]{mll_dist_ubarg_Lam2.png}
%\caption{Same as Fig.~\ref{fig:mlld_ist_Lam1} but with $\Lambda=2$ TeV.}
%\centering
%\label{fig:mlld_ist_Lam2}
%\end{figure}
%

\bibliographystyle{hunsrt.bst}
\bibliography{mybib2}

\end{document}